\documentclass[aps, pra, twocolumn, amsmath, amssymb, superscriptaddress, nofootinbib]{revtex4-1}

\makeatletter
\def\p@subsection{}
\def\p@subsubsection{}
\makeatother

\usepackage[utf8]{inputenc}
\usepackage{graphicx}
\usepackage{units}
\usepackage{color}
\usepackage[pdftex, colorlinks=true, linkcolor=myblue, citecolor=myblue,
urlcolor=myblue]{hyperref}
\usepackage{times}
\usepackage{comment}
\usepackage{graphicx}
\usepackage{amsmath, calc}
\usepackage{mathrsfs}
\usepackage{amsfonts}
\usepackage{amssymb}
\usepackage{bm}
\usepackage{bbm}
\usepackage{color}
\usepackage{array}
\usepackage{units}
\usepackage{tabstackengine}
\stackMath

\definecolor{myblue}{rgb}{0,0,1}
\definecolor{myred}{rgb}{1,0,0}

\newcommand{\bra}[1]{\langle #1|}
\newcommand{\ket}[1]{|#1\rangle}

\newcommand{\expect}[1]{\langle #1 \rangle}

\DeclareMathOperator{\Tr}{Tr}

\newmuskip\pFqmuskip
\newmuskip\tpFqmuskip

\newcommand*\pFq[6][8]{
  \begingroup 
  \pFqmuskip=#1mu\relax
  \mathchardef\normalcomma=\mathcode`,
  \mathcode`\,=\string"8000
  \begingroup\lccode`\~=`\,
  \lowercase{\endgroup\let~}\pFqcomma
  {}_{#2}F_{#3}{\left(\genfrac..{0pt}{}{#4}{#5};#6\right)}%
  \endgroup
}

\newcommand*\tpFq[6][8]{
  \begingroup 
  \tpFqmuskip=#1mu\relax
  \mathchardef\normalcomma=\mathcode`,
  \mathcode`\,=\string"8000
  \begingroup\lccode`\~=`\,
  \lowercase{\endgroup\let~}\pFqcomma
  {}_{#2}\tilde{F}_{#3}{\left(\genfrac..{0pt}{}{#4}{#5};#6\right)}%
  \endgroup
}

\newcommand{\pFqcomma}{{\normalcomma}\mskip\pFqmuskip}

\begin{document}


\title{A first encounter with exceptional points in a quantum model}


\author{C.~A.~Downing}
\email{c.a.downing@exeter.ac.uk} 
\affiliation{Department of Physics and Astronomy, University of Exeter, Exeter EX4 4QL, United Kingdom}

\author{V.~A.~Saroka}
\affiliation{Department of Physics, University of Rome Tor Vergata and INFN, Via della Ricerca Scientifica 1, 00133 Roma, Italy}
\affiliation{Institute for Nuclear Problems, Belarusian State University, Bobruiskaya 11, 220030 Minsk, Belarus}


\date{\today}


\begin{abstract}
\noindent
An exceptional point is a special point in parameter space at which two (or more) eigenvalues and eigenvectors coincide. The discovery of exceptional points within mechanical and optical systems has uncovered peculiar effects in their vicinity. Here we consider perhaps the simplest quantum model which exhibits an exceptional point and which allows for an analytical treatment. In particular, we re-examine a two-level atom driven by a laser and suffering from losses. The same exceptional point arises in several non-Hermitian matrices which determine various aspects of the dynamics of the system. There are consequences for some important observables, for example the spectrum evolves from being a Lorentzian-like singlet to a Mollow triplet upon passing through the exceptional point. Our analysis supports the perspective that viewing certain quantum systems through the lens of exceptional points offers some desirable explanatory advantages.
\end{abstract}


\maketitle



\section{Introduction}
\label{Sec:Introduction}


Modelling physical situations as eigenvalue problems requires physicists to frequently employ matrix diagonalization techniques on square matrices. All is well and good when dealing with diagonalizable matrices, but sometimes non-diagonalizable -- or defective -- matrices arise from seemingly reasonable physical theories~\cite{Berry2003, Bender2003}.

A defective $N \times N$ square matrix does not have $N$ linearly independent eigenvectors, and therefore it lacks a complete basis of eigenvectors. For example, consider the following pair of $2 \times 2$ matrices
\begin{equation}
\label{kgkgkgkhgkk:dfe}
M_1 =
\begin{pmatrix}
  a & 0   \\
  0 & b 
 \end{pmatrix},
 \quad\quad\quad
 M_2 =
\begin{pmatrix}
    a & 1   \\
  0 &   a 
 \end{pmatrix},
 \end{equation}
 where $a$ and $b$ are real numbers. Normal matrices, including Hermitian and unitary matrices, are not defective~\cite{Strang2006}. In the case of $M_1$, there are two distinct eigenvalues $a$ and $b$ which are associated with the linearly independent eigenvectors $\left(1, 0 \right)^\mathrm{T}$ and $\left(0, 1 \right)^\mathrm{T}$ respectively. However, the defective matrix $M_2$ has the repeated eigenvalue $a$, and only one eigenvector $\left(1, 0 \right)^\mathrm{T}$. When the diagonalization of some matrix $M$ is impossible, the best one can do is to arrive at the Jordan normal form $J = A^{-1} M A$, which features square blocks (instead of just the eigenvalues of $M$) along the diagonal of $J$. In general the procedure for finding the matrix $A$ is quite involved, as described in Ref.~\cite{Strang2006}. Such diagonalization problems typically arise in a course on linear algebra, where it is natural to connect the formal mathematical issue to real world examples such as those encountered within physics~\cite{Miri2019, Ozdemir2019}.
 
Defective matrices can arise in non-Hermitian physical systems~\cite{Ganainy2018, Downing2023, Fox2023}. Closed systems, which are isolated from the external environment, may be described using Hermitian operators which guarantee diagonalizability. However, non-Hermicity allows for the conservation of probability to be violated, which is typically the case in open systems which interact with their environment~\cite{Sabbaghzadeh2007, Bhattacharya2012}. In these situations energy or information may flow into or out of the system. This permits the possibility of a defective theory: by varying parameters in the governing non-Hermitian matrix one may find a point of non-diagonalizability -- an exceptional point -- at which (at least two) eigenvalues and eigenvectors coalesce simultaneously~\cite{Bender2007, Bender2023}. Accessible accounts of exceptional points within mechanical~\cite{Berntson2013, Dolfo2018}, electromagnetic~\cite{Xu2020, Lakhtakia2020} and acoustical~\cite{Lawrie2022} setups have already been provided. In particular, Ref.~\cite{Dolfo2018} presents an illuminating discussion of Newton’s classical equation for a damped harmonic oscillator from the standpoint of an exceptional point.

The response of classical systems in the neighbourhood of their exceptional points has traditionally led to surprising behaviors~\cite{Miri2019, Ozdemir2019}. More recently, the nature of exceptional points in quantum systems has been investigated experimentally, mostly with superconducting qubits~\cite{Naghiloo2019, Chen2022, Liang2023}. These open systems can be modelled with master equations~\cite{Sabbaghzadeh2007, Bhattacharya2012}, which can be manipulated to provide non-Hermitian matrices which govern the system's dynamics. We will pursue this theoretical framework here, which sits comfortably alongside a course encompassing quantum optics or related fields. Notably, exceptional points have been expertly introduced at a more general level in Ref.~\cite{Berry2003}, while excellent primers on the underlying topic of non-Hermitian physics are supplied by the foundational Refs.~\cite{Bender2007, Bender2023}.

In what follows we reconsider perhaps the simplest quantum system exhibiting an exceptional point: a two-level atom driven by a laser. This example arguably presents an ideal first encounter with spectral degeneracies within a quantum model. We discuss some remarkable phenomena occurring in the vicinity of the exceptional point. Most notably, the emission spectrum of the atom is reconstructed from being a simple singlet to an exotic triple-peaked structure after the system passes through its exceptional point, which is a particularly vivid manifestation of non-Hermitian physics.

The rest of this work is organized as follows. We start by laying out the closed system version of the model in Sec.~\ref{app:single_2sddsLS_model}. We then introduce the open system description in Sec.~\ref{app:single_2sdsdsddsLS_model}, which requires some manipulations with master equations at the level of Refs.~\cite{Sabbaghzadeh2007, Bhattacharya2012}. Armed with these theoretical foundations, we then reveal the exceptional points inherent to the model in Sec.~\ref{app:one_tisdxxzasdasdczxcasdadczxcsdme} and the differential equations necessary for studying the dynamics of the system in Sec.~\ref{app:ondfdfe_time}. The consequences for the population dynamics [Sec.~\ref{app:one_time}], first-order coherence [Sec.~\ref{app:one_tisdxczxcsdme}], second-order coherence [Sec.~\ref{app:one_tisdsdme}] and optical spectrum [Sec.~\ref{app:onsadsde_time}] are then discussed before we draw some conclusions in Sec.~\ref{Sec:Conclusion}.
\\


\section{Closed system}
\label{app:single_2sddsLS_model}

Let us consider an atom with a two-dimensional Hilbert space composed of the ground state $\ket{0}$ and the excited state $\ket{1}$. To move between these two states, we introduce the raising operator $\sigma^{\dagger} = \ket{1} \bra{0}$ and the lowering operator $\sigma = \ket{0} \bra{1}$, which satisfy the anticommutation relation $\{ \sigma, \sigma^{\dagger}\} = I$. These operators act so that $\sigma^{\dagger} \ket{0} = \ket{1}$ and $\sigma \ket{1} = \ket{0}$. If the transition frequency between the two levels is $\omega_0$, the corresponding energy can be described with the term $\hbar \omega_0 \ket{1} \bra{1}$. If the atom is coupled to a laser of amplitude $\Omega$, frequency $\omega_{\mathrm{L}}$ and phase $\theta$, the driving energy can be modelled with the term $\hbar \Omega \mathrm{e}^{\mathrm{i} \theta} \mathrm{e}^{-\mathrm{i}\omega_{\mathrm{L}} t} \ket{1} \bra{0}$ and its Hermitian conjugate. The Hamiltonian operator may then (in the rotating wave approximation) be written as~\cite{Merlin2020} 
\begin{equation}
\label{eq:apfghgp02}
 \hat{\mathcal{H}} = \hbar \omega_0 \sigma^{\dagger} \sigma + \hbar \Omega \mathrm{e}^{\mathrm{i}\theta} \mathrm{e}^{-\mathrm{i}\omega_{\mathrm{L}} t} \sigma^{\dagger} + \hbar \Omega \mathrm{e}^{-\mathrm{i}\theta} \mathrm{e}^{\mathrm{i}\omega_{\mathrm{L}} t} \sigma.
 \end{equation}
In order to remove the explicit time dependence in Eq.~\eqref{eq:apfghgp02}, we move into the rotating frame of the driving laser using the unitary transformation $U = \exp{ ( \mathrm{i} \omega_{\mathrm{L}} t \sigma^{\dagger} \sigma ) }$, which leads to
\begin{equation}
\label{eq:app02}
 \hat{H} = \Delta \sigma^{\dagger} \sigma + \Omega \mathrm{e}^{\mathrm{i}\theta} \sigma^{\dagger} + \Omega \mathrm{e}^{-\mathrm{i}\theta}\sigma,
 \end{equation}
 where we have set $\hbar = 1$ here and in all subsequent equations, and where the laser-atom detuning frequency $\Delta = \omega_0 - \omega_{\mathrm{L}}$. We also used the transformation equation $ \hat{H} = U  \hat{\mathcal{H}}  U^\dagger + \mathrm{i} ( \partial_t U ) U^\dagger $, along with the Baker–Campbell–Hausdorff formula $\mathrm{e}^A B \mathrm{e}^{-A} = B + [A, B] + \tfrac{1}{2} [A, [A, B]] + \cdots$ for two operators $A$ and $B$~\cite{Preble2013}. 
 
The Hamiltonian operator of Eq.~\eqref{eq:app02} is not diagonal in the $\ket{0}$ and $\ket{1}$ basis, so a suitable transformation needs to be found. We suppose that two new states $\ket{+}$ and $\ket{-}$ bring about the diagonal form of the Hamiltonian like so
\begin{equation}
\label{eq:appjhjsdsd2}
 \hat{H} = \omega_+ \ket{+} \bra{+} + \omega_- \ket{-} \bra{-}.
\end{equation}
As can be checked by direct substitution into Eq.~\eqref{eq:appjhjsdsd2}, the required superposition states are
\begin{align}
\label{eq:appjsddfdsdhjsdsd2}
 \ket{+} &= \mathrm{e}^{-\mathrm{i}\theta} \sin \phi \ket{0} + \cos \phi \ket{1},\\
  \ket{-} &=  \mathrm{e}^{-\mathrm{i}\theta} \cos \phi \ket{0} - \sin \phi \ket{1},  \label{eq:appjsddfdsdhvcvcjsdsd2} 
  \end{align}
where $\phi$ is defined through
\begin{equation}
\label{eq:appjsddsdsdsfdsdhjsdsd2}
 \sin \phi = \frac{1}{\sqrt{2}} \sqrt{ 1 - \frac{\Delta}{2 \nu } },
 \quad\quad
 \cos \phi = \frac{1}{\sqrt{2}} \sqrt{ 1 + \frac{\Delta}{2 \nu } }.
\end{equation}
The two energy levels, which are independent of the driving phase $\theta$, are given by
\begin{equation}
\label{eq:appjsdsdhjsdsd2}
 \omega_{\pm} = \frac{\Delta}{2} \pm \nu,
\end{equation}
where we have introduced the auxiliary frequency 
\begin{equation}
\label{eq:RRRRRRRRappjsdsdhjsdsd2}
  \nu = \sqrt{   \Omega^2 + \left(  \frac{\Delta}{2} \right)^2 }.
\end{equation}
Clearly there are no spectral degeneracies between the two states $\ket{+}$ and $\ket{-}$ in this completely closed system, and hence no exceptional points are present. This fact can also be seen by directly solving the Schrödinger equation $\mathrm{i} \partial_t \ket{ \psi } = \hat{H} \ket{ \psi }$ with the trial solution $\ket{ \psi } = A \ket{0} + B \ket{1}$. The dynamical matrix describing the system in the $\{ \ket{0}, \ket{1} \}$ basis is then
\begin{equation}
\label{sdffsdfdgfgdgffsd:dfe}
H =
\begin{pmatrix}
  0 & \Omega \mathrm{e}^{-\mathrm{i}\theta}   \\
  \Omega \mathrm{e}^{\mathrm{i}\theta} & \Delta   
 \end{pmatrix},
 \end{equation}
 whose eigenvalues are exactly those of Eq.~\eqref{eq:appjsdsdhjsdsd2}. Crucially, the governing matrix of Eq.~\eqref{sdffsdfdgfgdgffsd:dfe} is clearly seen to be Hermitian $H = H^\dagger$, forbidding any exceptional point physics. This fully Hermitian theory also ensures that the norm $\langle \psi | \psi \rangle = 1$ is both time-independent and probability preserving, as should be the case for a closed system.
\\


\section{Open system}
\label{app:single_2sdsdsddsLS_model}

In the style of Refs.~\cite{Sabbaghzadeh2007, Bhattacharya2012}, losses may be introduced into the model by upgrading the Hamiltonian dynamics of Sec.~\ref{app:single_2sddsLS_model}. At the more comprehensive level of the density matrix $\rho$, which accounts for mixed states and not just the pure states as described by some state vector $\ket{ \psi }$, the master equation reads~\cite{Manzano2020}
\begin{equation}
\label{eqapp:master}
 \partial_t \rho = \mathrm{i} [ \rho, \hat{H} ] +  \frac{\gamma}{2} \left( 2 \sigma \rho \sigma^{\dagger} -  \sigma^{\dagger} \sigma \rho - \rho \sigma^{\dagger} \sigma \right),
\end{equation}
where $\gamma \ge 0$ is the damping decay rate of the atom, and where the Lindblad form of this equation is derived in Ref.~\cite{Pearle2012}. The first term on the right-hand-side of Eq.~\eqref{eqapp:master} is the von Neumann equation and accounts for the dynamics of the closed system using the Hamiltonian operator from Eq.~\eqref{eq:app02}. Dissipation into the external environment is described by the second term on the right-hand-side of Eq.~\eqref{eqapp:master} which allows for non-unitary evolution~\cite{Sabbaghzadeh2007, Bhattacharya2012}. Since the atom is associated with the states $\ket{0}$ and $\ket{1}$ only, the density matrix is composed of just four elements
\begin{equation}
\label{sdffsdfssddsd:dfe}
\rho = \begin{pmatrix}
  \rho_{0, 0}  \\
  \rho_{1, 1} \\
    \rho_{1, 0} \\
      \rho_{0, 1} 
 \end{pmatrix},
 \end{equation}
 where $\rho_{n, m} = \langle n | \rho | m \rangle$ and $n, m \in \{ 0, 1 \}$. The necessary condition $\rho_{0, 0} + \rho_{1, 1} = 1$ ensures that probability is conserved. Using the master equation of Eq.~\eqref{eqapp:master} to compute the time evolution of the four matrix elements $\rho_{n, m}$ leads to the dynamical equation
 \begin{equation}
\label{sdfsdf:dfe}
\partial_t \rho = \mathcal{L} \rho,
 \end{equation}
where the $4 \times 4$ Liouvillian matrix
\begin{equation}
\label{sdffsdfsd:dfe}
 \mathcal{L} =
\begin{pmatrix}
  0 & \gamma & -\mathrm{i}\Omega \mathrm{e}^{-\mathrm{i}\theta} & \mathrm{i}\Omega \mathrm{e}^{\mathrm{i}\theta}  \\
  0 & -\gamma & \mathrm{i}\Omega \mathrm{e}^{-\mathrm{i}\theta} & -\mathrm{i}\Omega \mathrm{e}^{\mathrm{i}\theta}  \\
    -\mathrm{i}\Omega \mathrm{e}^{\mathrm{i}\theta} & \mathrm{i}\Omega \mathrm{e}^{\mathrm{i}\theta} & -\mathrm{i} \Delta -\frac{\gamma}{2} & 0  \\
      \mathrm{i}\Omega \mathrm{e}^{-\mathrm{i}\theta} & -\mathrm{i}\Omega \mathrm{e}^{-\mathrm{i}\theta} & 0 & \mathrm{i} \Delta -\frac{\gamma}{2} 
 \end{pmatrix},
 \end{equation}
 which is non-Hermitian $\mathcal{L} \ne \mathcal{L}^\dagger$, unlike the Hermitian matrix of Eq.~\eqref{sdffsdfdgfgdgffsd:dfe} which describes the closed version of the model. The non-Hermicity of Eq.~\eqref{sdffsdfsd:dfe} raises the possibility for exceptional point physics to arise thanks to the open nature of the system.
 
Before delving into the nature of the exceptional points and their consequences for the dynamics of the system, we lay the foundations for results needed later by considering the simplest response of the atom: its steady state behavior. Considering sufficiently large timescales $t \to +\infty$, such that the time derivative $\partial_t \rho = 0$, the steady state density matrix is given by the solution of the linear homogeneous equation $\mathcal{L} \rho = 0$. Since the determinant of Eq.~\eqref{sdffsdfsd:dfe} is zero, there are infinitely many solutions at first glance. Therefore, it is enough to solve just three of the simultaneous equations arising from the four-dimensional system $\mathcal{L} \rho = 0$ since the probability preservation condition $\rho_{0, 0} + \rho_{1, 1} = 1$ uniquely determines the density matrix. After some algebra, we obtain the steady state density matrix elements
 \begin{align}
\label{eqapxcvgfhgsdasdadsfdsvcvcfxxcvxp:umatrix}
\lim_{t \to +\infty}  \rho_{0, 0} &= \frac{ \Omega^2 + \Delta^2 + \left( \tfrac{\gamma}{2} \right)^2}{2 \Omega^2 + \Delta^2 + \left( \tfrac{\gamma}{2} \right)^2 },
\\
 \lim_{t \to +\infty}  \rho_{1, 1} &= \frac{ \Omega^2}{2 \Omega^2 + \Delta^2 + \left( \tfrac{\gamma}{2} \right)^2 }, \label{xcvxvcv:umatrix}
   \\
 \lim_{t \to +\infty}    \rho_{1, 0} = \lim_{t \to +\infty} \rho_{0, 1}^\ast &= \frac{ -\Omega \mathrm{e}^{\mathrm{i} \theta } \left( \Delta + \mathrm{i} \frac{\gamma}{2} \right) }{2 \Omega^2 + \Delta^2 + \left( \tfrac{\gamma}{2} \right)^2 }, \label{xcvxvcv:udfsfdsmatrix}
 \end{align} 
which will be employed later on. Having warmed up by manipulating the density matrix of the atom in the steady state, we now turn to searching for the exceptional points contained within the governing Liouvillian matrix.
  \\
 
\begin{figure*}[tb]
 \includegraphics[width=\linewidth]{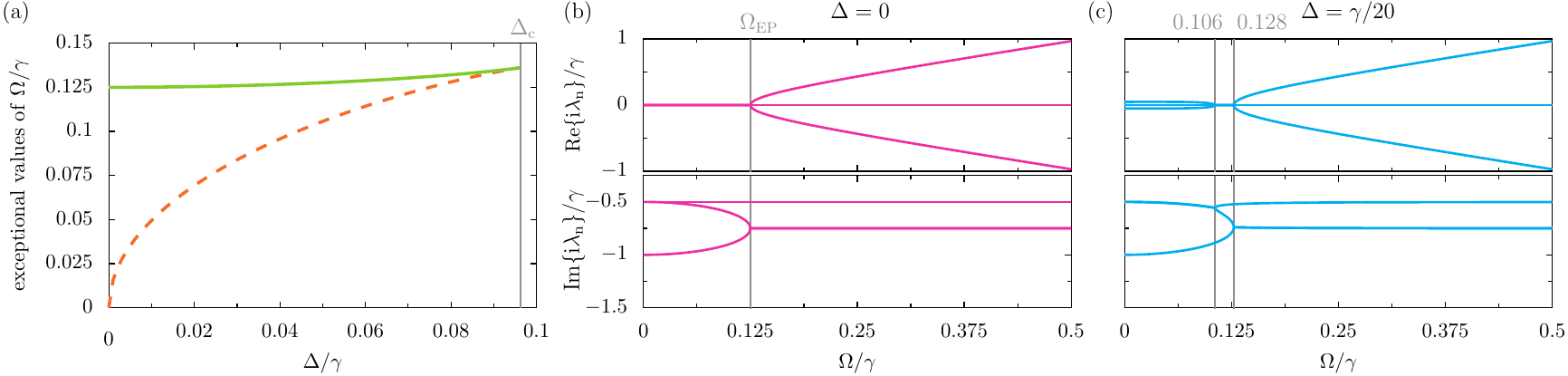}
 \caption{(Color online) \textbf{Exceptional points in a two-level atom.} (a): the values of the driving amplitude $\Omega$, in units of the damping rate $\gamma$, corresponding to an exceptional point in the Liouvillian matrix $\mathcal{L}$ [cf. Eq.~\eqref{sdffsdfsd:dfe}]. These values are found from the solutions of Eq.~\eqref{dfsfymbntrg}, which exist up to a critical detuning $\Delta_{\mathrm{c}}$ as marked by the vertical grey line [cf. Eq.~\eqref{eqapp:sdxcvxvxcvssddsdsd}]. (b): the real and imaginary parts of the three nontrivial Liouvillian eigenvalues $\mathrm{i} \lambda_1, \mathrm{i} \lambda_2$ and $\mathrm{i} \lambda_3$ as a function of $\Omega$ when $\Delta = 0$ [cf. Eq.~\eqref{eq:eiggsdfsdfg23}, Eq.~\eqref{eq:edfsdfbgfiggsdfsdfdsdfsffsffg23} and Eq.~\eqref{eqsdfsdfsdfn:eisdfsdfsggsdfsdfg23}]. Vertical grey line: exceptional point $\Omega_{\mathrm{EP}} = \gamma/8$. (c): as for panel (b), but for nonzero detuning where $\Delta = \gamma/20$ [cf. Eq.~\eqref{eq:eiggg23}, Eq.~\eqref{eqsadsda:eiggg23} and Eq.~\eqref{eq:eigggdfsfcsdvbr23}].}
 \label{exceppp}
\end{figure*}
 

\section{Exceptional points}
\label{app:one_tisdxxzasdasdczxcasdadczxcsdme}

The behaviour of the atom is fully described by the Hermitian matrix of Eq.~\eqref{sdffsdfdgfgdgffsd:dfe} when it is considered to be a lossless system. With the help of the non-Hermitian Liouvillian matrix of Eq.~\eqref{sdffsdfsd:dfe} dissipation is accounted for, which raises the chance for exceptional points to occur in the open version of the system.

The four complex eigenvalues $\lambda_ n$ of the Liouvillian matrix may be found from Eq.~\eqref{sdffsdfsd:dfe} via the quartic equation
   \begin{equation}
\label{cvdfgfgsdfsdfdxcvx}
 \lambda \left( \lambda^3 + a_2 \lambda^2 + a_1 \lambda  + a_0 \right) = 0, 
  \end{equation}
  where the three real-valued coefficients $a_n$ are given by
   \begin{align}
\label{eq:xcvxcvxcvxdxvdv}
a_ 2 &= 2 \gamma,  \\
a_ 1 &= \Delta^2 + 4 \Omega^2 + \frac{5}{4} \gamma^2,    \\
a_0 &= \gamma \left[ \Delta^2 + 2 \Omega^2 + \left( \tfrac{\gamma}{2} \right)^2 \right].  
\end{align}
The solution of the effective cubic equation within Eq.~\eqref{cvdfgfgsdfsdfdxcvx} defines the dynamics, while the fourth solution $\lambda_ 4 = 0$ corresponds to its steady state behaviour. The three nontrivial eigenvalues are
 \begin{align}
\lambda_ 1 &= -\frac{2 \gamma}{3} + S + T, \label{eq:eiggg23} \\
\lambda_ 2 &= -\frac{2 \gamma}{3} - \left( \frac{ S + T }{2}  \right) + \mathrm{i} \frac{\sqrt{3}}{2} \left( S - T \right),  \label{eqsadsda:eiggg23}  \\
\lambda_ 3 &= -\frac{2 \gamma}{3} -  \left( \frac{ S + T }{2}  \right) - \mathrm{i} \frac{\sqrt{3}}{2} \left( S - T \right),  \label{eq:eigggdfsfcsdvbr23}
\end{align}
where $S$ and $T$ are defined in terms of $D$ via
\begin{equation}
\label{eq:sdssdsdsdsdd}
S = \sqrt[3]{ R + \sqrt{D} },
~~~~~
T = \sqrt[3]{ R - \sqrt{D} },
~~~~~
 D = R^2 + Q^3.
\end{equation}
The three physical parameters of the problem (the detuning frequency $\Delta$, driving amplitude $\Omega$ and loss rate $\gamma$) enter the eigenvalues via the quantities 
 \begin{align}
\label{eq:egdfdgfdgfgdgiggsdfsdfg23}
R &= \frac{\gamma}{6} \left[ 2\Omega^2 - \Delta^2 - \left( \tfrac{\gamma}{6} \right)^2 \right],  \\
Q &= \frac{1}{3} \left[ 4\Omega^2 + \Delta^2 -  \tfrac{\gamma^2}{12}  \right]. 
\end{align}
Eigenvalue degeneracies occur when $\lambda_ 2 = \lambda_ 3$, so $S = T$. Equivalently $D = 0$, which leads to the sextic equation 
  \begin{equation}
\label{dfsfymbntrg}
 \Omega^6 + b_2 \Omega^4 + b_1 \Omega^2  + b_0 = 0, 
  \end{equation}
  where the three real-valued coefficients $b_n$ are given by
   \begin{align}
\label{eq:sgsegrdhrgdrgxcvxcvxcvxdxvdv}
b_ 2 &= 3 \left( \tfrac{\Delta}{2} \right)^2 - \left( \tfrac{\gamma}{8} \right)^2,  \\
b_ 1 &= \left( \tfrac{\Delta}{8} \right)^2 \left[ 12 \Delta^2 - 5 \gamma^2 \right],    \\
b_0 &= \left( \tfrac{\Delta}{8} \right)^2 \left[  \Delta^2 + \left( \tfrac{\gamma}{2} \right)^2 \right]^2. 
\end{align}
The discriminant $\mathcal{D}$ of the bi-cubic equation appearing within Eq.~\eqref{dfsfymbntrg} has the crucial proportionality
   \begin{equation}
\label{dfsfymbdsfsdfsfntrg}
 \mathcal{D} \propto \Delta^2 \gamma^4 \left[ \Delta^2 - \tfrac{1}{3} \left( \tfrac{\gamma}{6} \right)^2 \right]^3, 
  \end{equation}
  which determines the character of the bi-cubic solutions. Notably, the discriminant vanishes at a critical value of the detuning frequency
  \begin{equation}
\label{eqapp:sdxcvxvxcvssddsdsd}
\Delta_{\mathrm{c}} = \frac{\gamma}{6\sqrt{3}}.
\end{equation}
The exact cubic solutions of Eq.~\eqref{dfsfymbntrg} [not given here, but of the same formal structure as Eq.~\eqref{eq:eiggg23}, Eq.~\eqref{eqsadsda:eiggg23} and Eq.~\eqref{eq:eigggdfsfcsdvbr23}] allow one to find the values of $\Omega$ corresponding to eigenvalue degeneracies, which (when they arise along with eigenvector degneracies) are exceptional points. The results are plotted in Fig.~\ref{exceppp}~(a) as a function of $\Delta$, which shows that for nonzero $\Delta$ there are two exceptional points for a given detuning frequency (solid green and dashed orange lines) up to the critical detuning frequency $\Delta_{\mathrm{c}}$ (vertical grey line) above which no spectral degeneracies exist.

The preceding analysis is considerably simplified for the case of vanishing laser-atom detuning ($\Delta = 0$). In this resonant case the three eigenvalues $\lambda_n$ of Eq.~\eqref{eq:eiggg23}, Eq.~\eqref{eqsadsda:eiggg23} and Eq.~\eqref{eq:eigggdfsfcsdvbr23} reduce to
 \begin{align}
\lambda_ 1 &= -\frac{\gamma}{2}, \label{eq:eiggsdfsdfg23} \\
\lambda_ 2 &= -\frac{3\gamma}{4} + \mathrm{i} \tilde{\Omega}, \label{eq:edfsdfbgfiggsdfsdfdsdfsffsffg23}  \\
\lambda_ 3 &=  -\frac{3\gamma}{4} - \mathrm{i} \tilde{\Omega},   \label{eqsdfsdfsdfn:eisdfsdfsggsdfsdfg23}
\end{align}
where we have introduced the auxiliary frequency 
\begin{equation}
\label{eqapp:sdsdsd}
\tilde{\Omega} = \sqrt{ \left( 2 \Omega \right)^2 - \left( \tfrac{\gamma}{4} \right)^2 }.
\end{equation}
The real and imaginary parts of the complex eigenvalues $\mathrm{i} \lambda_n$ are plotted as the magenta lines in Fig.~\ref{exceppp}~(b), which showcases the presence of a single exceptional point (grey line) when the driving amplitude is equal to
\begin{equation}
\label{eqapp:sdssddsdsd}
\Omega_{\mathrm{EP}} = \frac{\gamma}{8}.
\end{equation}
 The unnormalized eigenvectors $\Lambda_{1,2, 3}$  corresponding to the eigenvalues $\lambda_{1,2, 3}$ are given by
 \begin{equation}
\label{eqapsdfsdfp:dgfgdbgbgbhgr}
\Lambda_{1} =
\begin{pmatrix}
  0   \\
 0  \\
  \mathrm{e}^{\mathrm{i} \theta}  \\
    \mathrm{e}^{-\mathrm{i} \theta}
 \end{pmatrix},
 ~
\Lambda_{2} =
\begin{pmatrix}
  \frac{-8 \mathrm{i} \Omega}{4 \mathrm{i} \tilde{\Omega} + \gamma}  \\
 \frac{8 \mathrm{i} \Omega}{ \gamma + 4 \mathrm{i} \tilde{\Omega} } \\
  \mathrm{e}^{\mathrm{i} \theta}  \\
   - \mathrm{e}^{-\mathrm{i} \theta}
 \end{pmatrix},
 ~
 \Lambda_{3} =
\begin{pmatrix}
  \frac{8 \mathrm{i} \Omega}{4 \mathrm{i} \tilde{\Omega} - \gamma}  \\
 \frac{8 \mathrm{i} \Omega}{\gamma-4 \mathrm{i} \tilde{\Omega}} \\
  \mathrm{e}^{\mathrm{i} \theta}  \\
   - \mathrm{e}^{-\mathrm{i} \theta}
 \end{pmatrix},
 \end{equation}
 which confirm the existence of an exceptional point at $\Omega_{\mathrm{EP}}$ due to the simultaneous coalescence of both the eigenvectors $\Lambda_{2}$ and $\Lambda_{3}$ and the eigenvalues $\lambda_{2}$ and $\lambda_{3}$. This analysis demonstrates the presence of an exceptional point in a simple quantum model.
 
 Let us now briefly return to the more complicated case of nonzero atom-laser detuning frequency, as is considered in the plots of Fig.~\ref{exceppp}~(c). For the example situation with $\Delta = \gamma/20$, we use the full eigenvalues $\lambda_n$ as defined through Eq.~\eqref{eq:eiggg23}, Eq.~\eqref{eqsadsda:eiggg23} and Eq.~\eqref{eq:eigggdfsfcsdvbr23} to plot the real and imaginary parts of $\mathrm{i}\lambda_n$ as the cyan lines in panel~(c). These plots demonstrate two exceptional points (vertical grey lines), as predicted by the locator plot provided in Fig.~\ref{exceppp}~(a).
 
 Armed with the knowledge of the locations of the exceptional points in the two-level atom, we are nearly ready to consider the consequences of exceptional point physics for some typical observables, but before that we briefly provide the equations for the atomic dynamics in Sec.~\ref{app:ondfdfe_time} which sets up the remaining calculations.
\\


\section{Atomic moments}
\label{app:ondfdfe_time}

 \begin{figure*}[tb]
 \includegraphics[width=\linewidth]{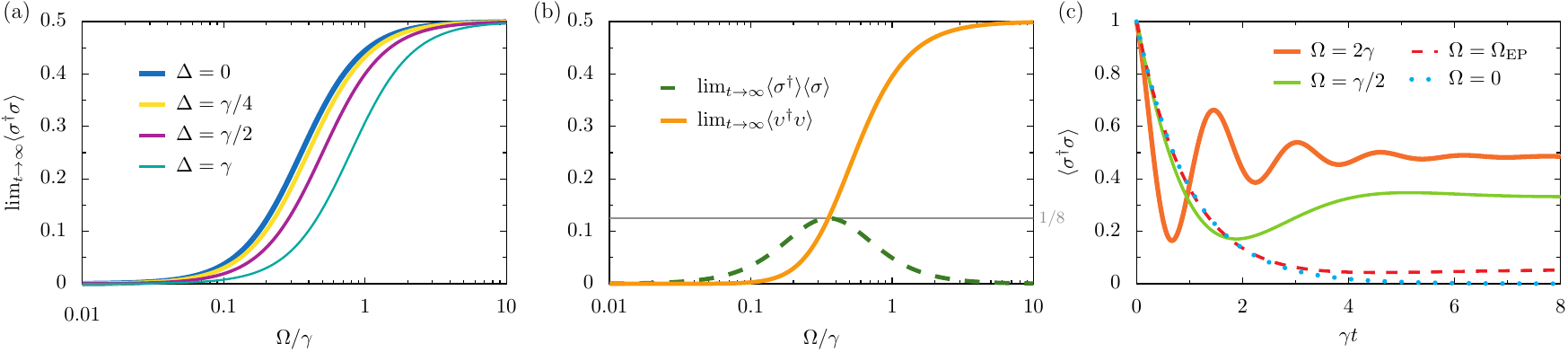}
 \caption{(Color online) \textbf{Population of a two-level atom.} (a): steady state population as a function of the driving amplitude $\Omega$, in units of the damping rate $\gamma$ [cf. Eq.~\eqref{eq:populationsd343434issi}]. (b): classical part (dashed green line) and quantum part (solid orange line) making up the steady state population when $\Delta = 0$ [cf. Eq.~\eqref{eq:appjsddfdfdfdsdhjsdsd2} and Eq.~\eqref{eq:appjsddfdfdfdsdhjsdsd22}]. Horizontal line: maximum of the classical field. (c): dynamic population of the atom as a function of time, in units of $\gamma^{-1}$ [cf. Eq.~\eqref{dsdfssdsdsddfs}, Eq.~\eqref{dsdfssdsdsddfs2} and Eq.~\eqref{dsdfssdsdsddfs3}]. Increasing values of $\Omega$ are displayed with increasingly thick lines, along with the result at $\Omega_{\mathrm{EP}}$ (dashed red line) and vanishing $\Omega$ (dotted cyan line). In panel (c), $\Delta = 0$ and the initial condition is $\langle \sigma^\dagger \sigma \rangle = 1$ at $t = 0$.}
 \label{popsingle}
\end{figure*}

The average value of an operator, for example the mean population of the atom as described by $\langle \sigma^{\dagger} \sigma \rangle$, can be obtained from the master equation of Eq.~\eqref{eqapp:master}. The relevant equation of motion arises after applying the trace property $\expect{\mathcal{O}} = \Tr{ \left( \mathcal{O} \rho \right) }$, which is valid for any operator $\mathcal{O}$, in the manner of Refs.~\cite{Sabbaghzadeh2007, Bhattacharya2012}. This process results in 
\begin{equation}
\label{eqapp:of_motion}
\mathrm{i }\partial_t \Psi  = \mathcal{H} \Psi + \mathcal{P},
\end{equation}
where the 3-vector of correlators $\Psi$ and the drive term $ \mathcal{P}$ are
\begin{equation}
\label{eqapp:umatrix}
\Psi =
\begin{pmatrix}
  \langle \sigma  \rangle   \\
  \langle {\sigma^{\dagger}} \rangle  \\
  \langle {\sigma^{\dagger}} \sigma \rangle 
 \end{pmatrix}, \quad\quad
 \mathcal{P} =
\begin{pmatrix}
\Omega  \mathrm{e}^{\mathrm{i}\theta}  \\
-\Omega  \mathrm{e}^{-\mathrm{i}\theta}   \\
    0 
 \end{pmatrix},
 \end{equation}
while the $3\times3$ dynamical matrix is given by 
\begin{equation}
\label{eqapp:m0matrix}
 \mathcal{H} = 
\begin{pmatrix}
 \Delta - \mathrm{i} \tfrac{\gamma}{2}  &  0 &  - 2  \Omega \mathrm{e}^{\mathrm{i}\theta} \\
   0 & -\Delta - \mathrm{i} \tfrac{\gamma}{2}   & 2 \Omega \mathrm{e}^{-\mathrm{i}\theta} \\
   - \Omega \mathrm{e}^{-\mathrm{i}\theta} &  \Omega \mathrm{e}^{\mathrm{i}\theta} & - \mathrm{i} \gamma
 \end{pmatrix}.
 \end{equation}
 The expectation values of the operators $\sigma$, $\sigma^\dagger$ and $\sigma^\dagger \sigma$ can then be found by solving the coupled equations of Eq.~\eqref{eqapp:of_motion} in the style of Ref.~\cite{Supplee2000} for example. The three complex eigenvalues of the dynamical matrix of Eq.~\eqref{eqapp:m0matrix} correspond to the three nonzero eigenvalues of the Liouvillian matrix of Eq.~\eqref{sdffsdfsd:dfe} (up to a factor of the imaginary unit $\mathrm{i}$), such that the locations of the exceptional points are wholly unaffected. For example, in the simplest case of resonance ($\Delta = 0$) one finds the three complex eigenvalues
  \begin{align}
\label{eq:eig43243gg23}
\epsilon_1 &= - \mathrm{i} \frac{\gamma}{2}, \\
\epsilon_2 &= - \mathrm{i} \frac{3\gamma}{4} - \tilde{\Omega}, \label{eq:eig43243gg23a} \\
\epsilon_3  &= - \mathrm{i} \frac{3\gamma}{4} + \tilde{\Omega},   \label{eq:eig43243gg23b}
\end{align}
where the frequency $\tilde{\Omega}$ is defined in Eq.~\eqref{eqapp:sdsdsd}, and where the exceptional point is given by Eq.~\eqref{eqapp:sdssddsdsd}.  A somewhat analogous treatment of the dynamics and exceptional point physics of a damped harmonic oscillator using purely classical equations is provided in Ref.~\cite{Dolfo2018}, which acts as a complement to the analysis presented here. In what follows we use the solutions of the coupled equations defined in Eq.~\eqref{eqapp:of_motion} to show how exceptional points can drastically change the dynamical response of the atom. One can expect some surprises due to changing character of Eq.~\eqref{eq:eig43243gg23a} and Eq.~\eqref{eq:eig43243gg23b} above and below $\Omega_{\mathrm{EP}}$ from complex quantities to wholly imaginary quantities. 
\\


\section{Atomic populations}
\label{app:one_time}

In the steady state, the average values for the operators contained within $\Psi$ [cf. Eq.~\eqref{eqapp:umatrix}] may then be found from $\Psi = - \mathcal{H}^{-1} \mathcal{P}$. The steady state population of the atom $\langle  \sigma^\dagger  \sigma \rangle$ and the steady state coherences $\langle  \sigma  \rangle$ and $\langle  \sigma^\dagger  \rangle$ then read [cf. Eq.~\eqref{eqapxcvgfhgsdasdadsfdsvcvcfxxcvxp:umatrix}, Eq.~\eqref{xcvxvcv:umatrix} and Eq.~\eqref{xcvxvcv:udfsfdsmatrix}] 
\begin{align}
\label{eq:populationsd343434issi}
\lim_{t \to +\infty} \langle \sigma^\dagger \sigma  \rangle &=  \lim_{t \to +\infty} \rho_{11},
\\
\lim_{t \to +\infty} \langle  \sigma  \rangle &= \lim_{t \to +\infty}  \rho_{10}, \label{eq:populationsd343434issi2}
\\
\lim_{t \to +\infty} \langle \sigma^\dagger   \rangle &=  \lim_{t \to +\infty} \rho_{01}. \label{eq:populationsd343434issi3}
\end{align}
The lower population bound of $\lim_{t \to +\infty} \langle \sigma^\dagger \sigma  \rangle  = 0$ is reached at vanishing driving ($\Omega \to 0$) since the atom resides in its ground state. The upper bound of $\lim_{t \to +\infty} \langle \sigma^\dagger \sigma  \rangle = 1/2$ is met in the large driving limit ($\Omega \to +\infty$) which reveals that population inversion is prohibited in this system. We plot the steady state atom population in Fig.~\ref{popsingle}~(a) as a function of the driving amplitude $\Omega$, which shows the evolution of the steady state population between the two bounds $0$ and $1/2$. A smaller $\Delta$ implies higher a population since driving close to resonance excites the atom most efficiently.

The underlying nature of the average atom population $\langle \sigma^\dagger \sigma  \rangle$ can be probed by decomposing the operator $\sigma$ into two parts $\sigma = \langle \sigma  \rangle + \upsilon$. Here the mean-field quantity $\langle \sigma  \rangle$ is a complex number accounting for the classical behaviour, while the operator $\upsilon$ tracks the quantum field. The operator $\sigma^\dagger \sigma$ can then be decomposed as $\sigma^\dagger \sigma =  \langle \sigma^\dagger  \rangle \langle \sigma  \rangle + \upsilon^\dagger \upsilon + \langle \sigma^\dagger  \rangle \upsilon + \langle \sigma  \rangle \upsilon^\dagger$. Since the quantum field $\upsilon$ has no mean field by construction $\langle \upsilon  \rangle = 0$ and the mean population has no interference terms
\begin{equation}
\label{eq:popuasdaddfdfdfsdsddfdfdflationsd343434issi}
\langle \sigma^\dagger \sigma \rangle =  \langle \sigma^\dagger  \rangle \langle \sigma  \rangle + \langle \upsilon^\dagger \upsilon \rangle.
\end{equation}
Therefore the classical and quantum contributions to the mean population in the steady state can be deciphered as
\begin{align}
\label{eq:appjsddfdfdfdsdhjsdsd2}
\lim_{t \to +\infty} \langle \sigma^\dagger  \rangle \langle \sigma  \rangle &= \frac{ \Omega^2 \left( \Delta^2 + \left( \tfrac{\gamma}{2} \right)^2 \right) }{\left( 2\Omega^2 + \Delta^2 + \left( \tfrac{\gamma}{2} \right)^2 \right)^2},\\
 \lim_{t \to +\infty} \langle \upsilon^\dagger \upsilon \rangle &= \frac{ 2\Omega^4 }{\left( 2\Omega^2 + \Delta^2 + \left( \tfrac{\gamma}{2} \right)^2 \right)^2}. \label{eq:appjsddfdfdfdsdhjsdsd22}
\end{align}
We plot both contributions to Eq.~\eqref{eq:popuasdaddfdfdfsdsddfdfdflationsd343434issi} as a function of $\Omega$ in Fig.~\ref{popsingle}~(b) for the simplest case of zero detuning. The classical field population described by the dashed green line and Eq.~\eqref{eq:appjsddfdfdfdsdhjsdsd2} has the lower bound of $0$ (occurring with both weak and strong driving) and an upper bound of just $1/8$, as is marked by the thin grey line. The quantum field associated with Eq.~\eqref{eq:appjsddfdfdfdsdhjsdsd22} varies more widely between $0$ and $1/2$ (solid orange line) and starts to dominate above $\Omega = \gamma/(2\sqrt{2})$. It is the only contribution in the large driving limit $\Omega \gg \gamma$, highlighting the quantum nature of the atom at saturation.

The full dynamical population of the atom follows from the time-dependent solution of Eq.~\eqref{eqapp:of_motion}. With zero detuning and with the initial population $\langle {\sigma^{\dagger}} \sigma \rangle = 1$ at $t=0$ the expressions for the atom populations are considerably simplified. Taking into account the exceptional point at $\Omega_{\mathrm{EP}}$ we find that the non-decaying part of the solution takes the form of trigonometric oscillatory functions when the driving is strong ($\Omega > \Omega_{\mathrm{EP}}$), hyperbolic functions when the driving is weak ($\Omega < \Omega_{\mathrm{EP}}$), and otherwise a linear in time function exactly at the exceptional point ($\Omega = \Omega_{\mathrm{EP}}$):
\begin{widetext} 
\begin{align}
\label{dsdfssdsdsddfs}
\langle \sigma^\dagger \sigma  \rangle &=  \frac{\Omega^2}{2\Omega^2+\left( \frac{\gamma}{2} \right)^2} + \frac{ 1 }{2\Omega^2+\left( \frac{\gamma}{2} \right)^2} \biggl\{ \left[ \Omega^2+\left( \tfrac{\gamma}{2} \right)^2 \right] \cos \left( \tilde{\Omega} t \right) - \frac{\gamma}{4 \tilde{\Omega} } \left[ 5 \Omega^2+\left( \tfrac{\gamma}{2} \right)^2 \right] \sin \left( \tilde{\Omega} t \right)  \biggl\} \mathrm{e}^{- \frac{3 \gamma t}{4}  },  &&(\Omega > \Omega_{\mathrm{EP}}),  \\
\langle \sigma^\dagger \sigma  \rangle &= \frac{1}{18} + \frac{1}{3} \left( \frac{17}{6} - \frac{7}{8} \gamma t \right) \mathrm{e}^{- \frac{3 \gamma t}{4}  }, &&(\Omega = \Omega_{\mathrm{EP}}), \label{dsdfssdsdsddfs2}\\
\langle \sigma^\dagger \sigma  \rangle & = \frac{\Omega^2}{2\Omega^2+\left( \frac{\gamma}{2} \right)^2} + \frac{ 1 }{2\Omega^2+\left( \frac{\gamma}{2} \right)^2} \biggl\{ \left[ \Omega^2+\left( \tfrac{\gamma}{2} \right)^2 \right] \cosh \left( \Gamma t \right) - \frac{\gamma}{4 \Gamma} \left[ 5 \Omega^2+\left( \tfrac{\gamma}{2} \right)^2 \right] \sinh \left( \Gamma t \right)  \biggl\} \mathrm{e}^{- \frac{3 \gamma t}{4}  }, &&(\Omega < \Omega_{\mathrm{EP}}), \label{dsdfssdsdsddfs3}
\end{align}
\end{widetext} 
where $\tilde{\Omega}$ is defined in Eq.~\eqref{eqapp:sdsdsd} and where
\begin{equation}
\label{eq:dfgdgvdv}
\Gamma = \sqrt{ \left( \tfrac{\gamma}{4} \right)^2 - \left( 2 \Omega \right)^2 }.
\end{equation}
We plot the dynamical atom population in Fig.~\ref{popsingle}~(c) as a function of time. Strikingly, only driving amplitudes above the exceptional point lead to the onset of population cycles (medium green line) which are increasingly pronounced for larger driving amplitudes (thick orange line). Therefore, within this system, by evolving from non-oscillatory to oscillatory atomic population dynamics one can infer that an exceptional point (dashed red line) has been safely passed through. Within classical mechanics, a similar dynamical transition is seen to occur for a damped harmonic oscillator moving between its under-damped and over-damped regimes~\cite{Dolfo2018}.

In the two limiting cases of vanishing driving and very strong driving 
\begin{align}
\label{eq:dfgdgdsfdsfdvdv}
\lim_{\Omega \to 0} \langle \sigma^\dagger \sigma  \rangle &= \mathrm{e}^{- \gamma t}, \\
\lim_{\Omega \to +\infty} \langle \sigma^\dagger \sigma  \rangle &=\frac{1}{2} \left( 1 + \cos \left( 2 \Omega t \right) \mathrm{e}^{-\frac{3\gamma t}{4}} \right). \label{eq:dfgdgdsfsfgvefth66dsfdvdv}
\end{align}
The purely exponential decay (with the time constant $1/\gamma$) of the undriven atom is seen as with the dotted cyan line in Fig.~\ref{popsingle}~(c). The very strong driving result of Eq.~\eqref{eq:dfgdgdsfsfgvefth66dsfdvdv} is not shown, but the expression reveals cosinusoidal oscillations (maximally bound between $1$ and $0$) which are exponentially damped towards a steady state population of $1/2$ with the characteristic time constant of $4/3\gamma$.
\\

\begin{figure*}[tb]
 \includegraphics[width=1.0\linewidth]{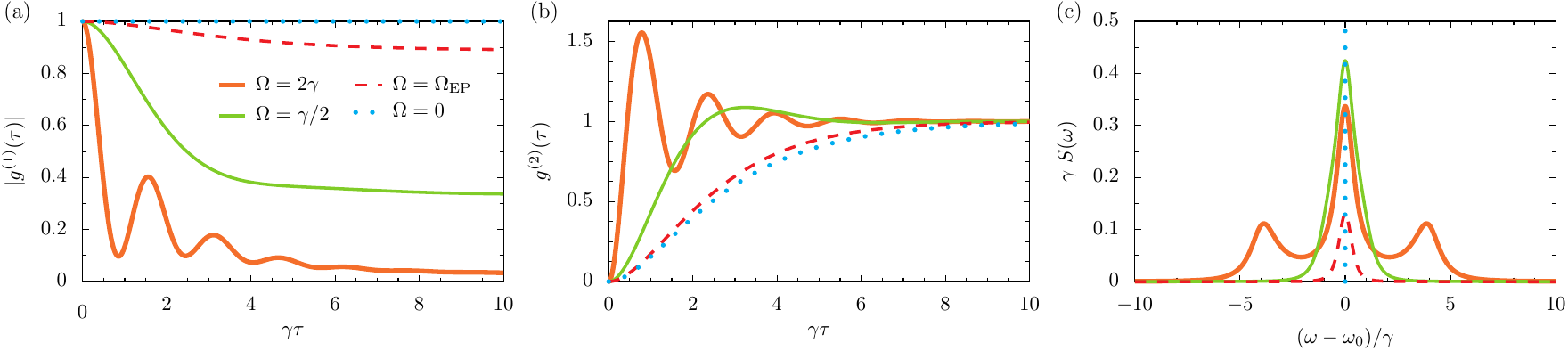}
 \caption{(Color online) \textbf{Degrees of coherence and optical spectrum of an atom.} (a): absolute value of the first-order coherence $g^{(1)} (\tau)$ as a function of the time delay $\tau$, in units of the inverse damping rate $\gamma^{-1}$ [cf. Eq.~\eqref{geetwo3ghjghj2sdsds3}, Eq.~\eqref{geetwo3ghjghj2sdsds33} and Eq.~\eqref{geetwo3ghjghj2sdsds333}]. (b): second-order coherence $ g^{(2)} (\tau)$ as a function of $\tau$ [cf. Eq.~\eqref{geetwo32sdsds3}, Eq.~\eqref{geetwo32sdsds3a} and Eq.~\eqref{geetwo32sdsds3b}]. (c): optical spectrum $S( \omega )$ as a function of the frequency $\omega$ of the emitted photons [cf. Eq.~\eqref{opty23sdsdsdads23}, Eq.~\eqref{opty23sdsdsdads23a} and Eq.~\eqref{opty23sdsdsdads23b}]. In this figure we consider $\Delta = 0$ and present several values of the driving amplitude $\Omega$ as marked by the legend in panel (a), as well as the results at the exceptional point (dashed red lines) and vanishing $\Omega$ (dotted cyan lines).}
 \label{geeone}
\end{figure*}


\section{First-order coherence}
\label{app:one_tisdxczxcsdme}

The temporal stability of the light emitted from the atom can be described by its coherence in time, as judged by correlation functions~\cite{Basano1982, Kuusela2018}. In particular, the first-order degree of coherence $g^{(1)} (\tau)$ is concerned with the first power of the light field, while the second-order degree of coherence $ g^{(2)} (\tau) $ is associated with the second power of the light field. Several neat experiments to measure the degrees of coherence of photons have been proposed~\cite{Kuusela2017, Scholz2018, Ferreira2020}, which allows one to characterize the nature of the emitted light. There are three kinds of first-order coherence, which allows one to determine the degree to which the light is monochromatic or not~\cite{Gardiner2014}. Fully coherent (and perfectly monochromatic) light occurs when $|g^{(1)} (\tau)| = 1$, light with partial coherence manifests itself when $0 < |g^{(1)} (\tau)| < 1$, and perfectly incoherent light arises when $|g^{(1)} (\tau)| = 0$. For example, a reasonable approximation to full coherence is the monochromatic light emitted from a laser, while a good approximation to full incoherence is the white light emanating from a thermal source. The degree of first-order coherence $ g^{(1)} (\tau) $ is defined via~\cite{Gardiner2014}
\begin{equation}
\label{geetwzxczcxo323}
 g^{(1)} (\tau) = \lim_{t\to +\infty} \frac{\langle \sigma^\dagger (t)  \sigma (t+\tau)  \rangle}{\langle \sigma^\dagger (t) \sigma (t) \rangle},
\end{equation}
where the time delay $\tau \ge 0$. The master equation of Eq.~\eqref{eqapp:master} yields the necessary equation of motion for the set of two-time correlators~\cite{Gardiner2014}
\begin{equation}
\label{eqapp:of_msdshkjhkdsdotion}
\mathrm{i} \partial_{\tau} \psi (t, t+\tau)  = \mathcal{P} \langle \sigma^\dagger (t) \rangle + \mathcal{H} \psi (t, t+\tau),
\end{equation}
\begin{equation}
\label{eqapp:umsfdfshkjkhefsdfatrix}
\psi (t, t+\tau)  =
\begin{pmatrix}
  \langle \sigma^{\dagger} (t) \sigma (t+\tau)   \rangle   \\
  \langle \sigma^{\dagger} (t) \sigma^{\dagger} ( t+\tau)  \rangle  \\
  \langle \sigma^{\dagger} (t) \sigma^{\dagger} (t+\tau) \sigma (t+\tau) \rangle 
 \end{pmatrix},
 \end{equation}
where $\mathcal{P}$ and $\mathcal{H}$ are defined in Eq.~\eqref{eqapp:umatrix} and Eq.~\eqref{eqapp:m0matrix} respectively. The degree of first-order coherence follows from the solution of Eq.~\eqref{eqapp:of_msdshkjhkdsdotion}, subject to the steady state normalization given by the denominator in Eq.~\eqref{geetwzxczcxo323}. We consider zero detuning so that the resulting analytic expressions are more compact. Depending on the value of $\Omega$, we have the absolute values:
\begin{widetext} 
\begin{align}
\label{geetwo3ghjghj2sdsds3}
 |g^{(1)}(\tau)| &= \frac{ \left( \tfrac{\gamma}{2} \right)^2}{2\Omega^2+\left( \tfrac{\gamma}{2} \right)^2} + \frac{\mathrm{e}^{-\frac{\gamma \tau}{2}  }}{2} + \left[ \left( \Omega^2-\tfrac{\gamma^2}{8} \right) \cos \left( \tilde{\Omega} \tau \right) + \frac{10\Omega^2-\left( \tfrac{\gamma}{2} \right)^2}{8 \tilde{\Omega} / \gamma} \sin \left( \tilde{\Omega} \tau  \right) \right] \frac{\mathrm{e}^{-\frac{3 \gamma \tau }{4} }}{2\Omega^2+\left( \tfrac{\gamma}{2} \right)^2}, &&(\Omega > \Omega_{\mathrm{EP}}), \\
  |g^{(1)}(\tau)| &= \frac{8}{9} + \frac{\mathrm{e}^{-\frac{\gamma \tau}{2}  }}{2} - \left(  \frac{ 28 + 3 \gamma \tau }{72} \right) \mathrm{e}^{-\frac{3 \gamma \tau }{4} },  &&(\Omega = \Omega_{\mathrm{EP}}),  \label{geetwo3ghjghj2sdsds33} \\
   |g^{(1)}(\tau)| &= \frac{ \left( \tfrac{\gamma}{2} \right)^2}{2\Omega^2+\left( \tfrac{\gamma}{2} \right)^2} + \frac{\mathrm{e}^{-\frac{\gamma \tau}{2}  }}{2} + \left[ \left( \Omega^2-\tfrac{\gamma^2}{8} \right) \cosh \left( \Gamma \tau \right) + \frac{10\Omega^2-\left( \tfrac{\gamma}{2} \right)^2}{8 \Gamma / \gamma} \sinh \left( \Gamma \tau  \right) \right] \frac{\mathrm{e}^{-\frac{3 \gamma \tau }{4}  }}{2\Omega^2+\left( \tfrac{\gamma}{2} \right)^2}, &&(\Omega < \Omega_{\mathrm{EP}}), \label{geetwo3ghjghj2sdsds333}
\end{align}
\end{widetext} 
where $\tilde{\Omega}$ and $\Gamma$ are defined in Eq.~\eqref{eqapp:sdsdsd} and Eq.~\eqref{eq:dfgdgvdv} respectively. The asymptotic behaviors of $|g^{(1)}(\tau)|$ at short and long time delays are captured by the expressions
\begin{align}
\label{dfgmnguuuydfg}
\lim_{\tau \to 0} |g^{(1)}(\tau)| &= 1, \\
\lim_{\tau \to +\infty} |g^{(1)}(\tau)| &= \frac{ \left( \tfrac{\gamma}{2} \right)^2}{2\Omega^2+\left( \tfrac{\gamma}{2} \right)^2}. \label{dfgmnguusdaduydfg}
\end{align}
At zero time delay the correlator is exactly unity, representing full coherence, which comes directly from the definition of Eq.~\eqref{geetwzxczcxo323}. With large time delays there is a greater chance of randomness and so the correlator reduces to a small and constant amount [cf. Eq.~\eqref{dfgmnguusdaduydfg}]. In between these extreme temporal delays, the exceptional point at $\Omega_{\mathrm{EP}}$ defines the crossover between regimes of either oscillating or non-oscillating coherences [cf. Eq.~\eqref{geetwo3ghjghj2sdsds3} and Eq.~\eqref{geetwo3ghjghj2sdsds333}], similar to the population dynamics in the previous section. In the vanishing driving and very strong driving limits the first-order coherence function reduces to
\begin{align}
\label{dfgmngudcsduuydfg}
\lim_{\Omega \to 0} |g^{(1)}(\tau)| &= 1, \\
\lim_{\Omega \to +\infty} |g^{(1)}(\tau)| &= \frac{1}{2} \left( 1 + \cos \left( 2 \Omega t \right) \mathrm{e}^{-\frac{\gamma \tau}{4}} \right) \mathrm{e}^{-\frac{\gamma \tau}{2}}. \label{dfgmngudcsduuydfdsfsdfsghghrgg}
\end{align}
The weak driving expression describes perfect monochromatic light due to the coherent driving of the laser, as represented by the cyan line in Fig.~\ref{geeone}~(a). The very strong driving result of Eq.~\eqref{dfgmngudcsduuydfdsfsdfsghghrgg} is not shown in panel (a) but exhibits cosinusoidal oscillations between $0$ and $1$ which are quickly washed out to zero with longer delay times.
\\


\section{Second-order coherence}
\label{app:one_tisdsdme}

Intensity correlations of the atom can be captured by the degree of second-order coherence $g^{(2)} (\tau)$, which can be readily measured experimentally with photon counting setups~\cite{Kuusela2017, Scholz2018, Ferreira2020}. This second-order correlator may be defined by~\cite{Gardiner2014}
\begin{equation}
\label{geetwo323}
 g^{(2)} (\tau) = \lim_{t\to +\infty} \frac{\langle \sigma^\dagger (t) \sigma^\dagger (t+\tau) \sigma (t+\tau)  \sigma (t) \rangle}{\langle \sigma^\dagger (t) \sigma (t) \rangle^2},
\end{equation}
for a time delay $\tau \ge 0$ between two different emissions, and with the normalization taken in the steady state. This important quantity describes a classical bunching effect when $g^{(2)} (0) >  g^{(2)} (\tau)$, where photon emissions are clumped together in time. Conversely, antibunching occurs when $g^{(2)} (0) <  g^{(2)} (\tau)$. This is a decidedly non-classical phenomena where the emitted photons are well separated in time from one another~\cite{Gardiner2014}. The edge case of $g^{(2)} (0) =  g^{(2)} (\tau)$ suggests photons are emitted with a random spacing in time. The master equation of Eq.~\eqref{eqapp:master} yields the necessary equation of motion for the set of two-time correlators~\cite{Gardiner2014}
\begin{equation}
\label{eqapp:of_msdsdsdotion}
\mathrm{i} \partial_{\tau} \Phi (t, t+\tau)  = \mathcal{P} \langle \sigma^\dagger (t) \sigma (t) \rangle + \mathcal{H} \Phi (t, t+\tau), 
\end{equation}
\begin{equation}
\label{eqapp:umsfdfsefsdfatrix}
\Phi (t, t+\tau)  =
\begin{pmatrix}
  \langle \sigma^{\dagger} (t) \sigma (t+\tau) \sigma (t)  \rangle   \\
  \langle \sigma^{\dagger} (t) \sigma^{\dagger} ( t+\tau) \sigma (t) \rangle  \\
  \langle \sigma^{\dagger} (t) \sigma^{\dagger} (t+\tau) \sigma (t+\tau) \sigma (t) \rangle 
 \end{pmatrix},
 \end{equation}
where $\mathcal{P}$ and $\mathcal{H}$ are defined in Eq.~\eqref{eqapp:umatrix} and Eq.~\eqref{eqapp:m0matrix} respectively. The degree of second-order coherence follows from the solution of Eq.~\eqref{eqapp:of_msdsdsdotion} subject to the steady state normalization of Eq.~\eqref{geetwo323}. We consider zero detuning so that the resulting analytic expressions are more manageable, and list the solutions for the regions $\Omega > \Omega_{\mathrm{EP}}$, $\Omega = \Omega_{\mathrm{EP}}$ and $\Omega < \Omega_{\mathrm{EP}}$ in that order:
\begin{align}
\label{geetwo32sdsds3}
 g^{(2)}(\tau) &= 1 - \left[ \cos \left( \tilde{\Omega} \tau  \right) + \frac{3\gamma}{4 \tilde{\Omega}} \sin \left( \tilde{\Omega} \tau \right) \right] \mathrm{e}^{-\frac{3\gamma \tau}{4}} ,   \\
  g^{(2)}(\tau) &= 1 - \left( 1 + \frac{3 \gamma \tau}{4}  \right) \mathrm{e}^{-\frac{3\gamma \tau}{4} }, \label{geetwo32sdsds3a} \\
   g^{(2)}(\tau) &= 1 - \left[ \cosh \left( \Gamma \tau  \right) + \frac{3\gamma}{4 \Gamma} \sinh \left( \Gamma \tau  \right) \right] \mathrm{e}^{-\frac{3\gamma \tau}{4} }, \label{geetwo32sdsds3b}
\end{align}
where $\tilde{\Omega}$ and $\Gamma$ are given by Eq.~\eqref{eqapp:sdsdsd} and Eq.~\eqref{eq:dfgdgvdv} respectively. The asymptotics of the full solutions at short and long delay times are
\begin{align}
\label{dfgmnguusdfsdfuydfg}
\lim_{\tau \to 0} g^{(2)}(\tau) &= 0, \\
\lim_{\tau \to +\infty} g^{(2)}(\tau) &= 1. \label{dfgmnguusdfdfssfsdfuydfg}
\end{align}
The vanishing time delay result of zero describes non-classical, antibunching behavior. This perfect antibunching follows since the two-level atom cannot simultaneously emit two photons: a finite delay time is necessary for the atom to be re-excited from its ground state in order to re-emit another photon. The opposing limit implies randomly bunched light, which arises because for large time delays the field intensities at $t$ and $t + \tau$ should be completely uncorrelated. Otherwise, for intermediate delay times, the full solutions of Eq.~\eqref{geetwo32sdsds3}, Eq.~\eqref{geetwo32sdsds3a} and Eq.~\eqref{geetwo32sdsds3b} showcase the remarkable impact of the exceptional point residing at $\Omega_{\mathrm{EP}}$. The correlator $g^{(2)}(\tau)$ either displays characteristic oscillations -- or not -- depending upon whether the driving amplitude has passed through the exceptional point $\Omega_{\mathrm{EP}}$ (Fig.~\ref{geeone}~(b)) since the nature of the eigenvalues change from being complex to wholly imaginary. xfIn the two limiting cases of vanishingly weak driving and very strong driving, we find the exact expressions reduce to
\begin{align}
\label{eqapp:of_msdsdfdfsfsdsdfsdfdsdotion}
 \lim_{\Omega \to 0} g^{(2)}(\tau) &= \left( 1 - \mathrm{e}^{-\frac{\gamma \tau}{2}}  \right)^2, \\
  \lim_{\Omega \to +\infty} g^{(2)}(\tau) &= 1 - \cos \left( 2 \Omega \tau \right) \mathrm{e}^{-\frac{3\gamma \tau}{4}}. \label{eqapp:of_msdsdfdfsfsdsdfsdfdsdotionBB}
\end{align}
The weak driving result is denoted by the cyan line in Fig.~\ref{geeone}~(b), and its lack of oscillations as it develops from $0$ to $1$ monotonically in time is a distinguishing feature. The strong driving result is not shown in panel (b), but the highly oscillatory expression displays extrema at $0$ and $2$, which will eventually damp out to unity with the time constant $4/3 \gamma$.
\\


\section{Optical spectrum}
\label{app:onsadsde_time}

The optical spectrum of the fluorescent light emitted by the atom $S(\omega)$ is defined in terms of the Fourier transform of the two-time correlator $\langle \sigma^{\dagger} (t) \sigma ( t+\tau)  \rangle$. The spectrum measures the intensity of the photons emitted by the atom at the frequency $\omega$, and mathematically it reads~\cite{Gardiner2014}
\begin{equation}
\label{opty2323}
 S (\omega) =  \lim_{t\to +\infty} \frac{1}{\pi} \frac{\mathrm{Re} \int_0^\infty  \langle \sigma^{\dagger} (t) \sigma ( t+\tau)  \rangle \mathrm{e}^{\mathrm{i} \omega \tau} \mathrm{d}\tau }{\langle \sigma^{\dagger} (t) \sigma ( t)  \rangle},
\end{equation}
which has been normalized so that $\int_0^\infty S (\omega) \mathrm{d}\omega= 1$. This ensures that $S (\omega)$ can be treated as the probability of the system emitting a photon at the frequency $\omega$. By inserting the solutions already found in Sec.~\ref{app:one_tisdxczxcsdme} for the first-order correlator $g^{(1)} (\tau)$ into the integrand of Eq.~\eqref{opty2323}, and after carrying out the time integral, the explicit form of the spectral lineshape may be found~\cite{Mollow1969}. With zero atom-laser detuning the optical spectrum may be written as
\begin{widetext} 
\begin{align}
\label{opty23sdsdsdads23}
S( \omega ) &=   \frac{ \left( \tfrac{\gamma}{2} \right)^2}{\left( \tfrac{\gamma}{2} \right)^2 + 2\Omega^2} \delta (\omega - \omega_0) + \frac{1}{2 \pi} \frac{ \tfrac{\gamma}{2} }{ ( \tfrac{\gamma}{2} )^2 + \left( \omega - \omega_0 \right) ^2 }  +  \sum_{\tau = \pm 1} S_\tau^{>} (\omega),  &&(\Omega > \Omega_{\mathrm{EP}}), \\
 S( \omega ) &=  \frac{8}{9} \delta (\omega - \omega_0) + \frac{1}{2 \pi} \frac{ \tfrac{\gamma}{2} }{ ( \tfrac{\gamma}{2} )^2 + \left( \omega - \omega_0 \right) ^2 } - \frac{\gamma}{4\pi} \frac{ \left( \omega - \omega_0 \right) ^2  + 3 \left( \frac{\gamma}{2} \right)^2 }{ \left[ \left( \omega - \omega_0 \right) ^2 + \left( \frac{3\gamma}{4} \right)^2 \right]^2 },  &&(\Omega = \Omega_{\mathrm{EP}}) \label{opty23sdsdsdads23a}, \\
 S( \omega ) &=   \frac{ \left( \tfrac{\gamma}{2} \right)^2}{\left( \tfrac{\gamma}{2} \right)^2 + 2\Omega^2} \delta (\omega - \omega_0) + \frac{1}{2 \pi} \frac{ \tfrac{\gamma}{2} }{ ( \tfrac{\gamma}{2} )^2 + \left( \omega - \omega_0 \right) ^2 }  + \sum_{\tau = \pm 1} \frac{L_\tau^{<}}{\pi} \frac{\frac{3\gamma}{4}-\tau \Gamma}{ \left( \frac{3\gamma}{4} - \tau \Gamma \right)^2 + \left( \omega - \omega_0 \right) ^2} , &&(\Omega < \Omega_{\mathrm{EP}}) \label{opty23sdsdsdads23b},
\end{align}
\end{widetext} 
where $\delta (x)$ is Dirac's delta function. The first term, describing a weighted delta spectral peak or `Rayleigh scattering peak', is common to all three expressions given above, and arises from the light elastically scattered by the atom from the driving laser~\cite{Mollow1969}. The second term is a standard Lorentzian term which emerges due to the relaxation process from the excited state to the ground state and it defines the central (and unshifted) spectral peak, with the characteristic broadening $\gamma$. Finally, the third term accounts for the spectral differences due to the size of the driving amplitude, including the possible presence of sidebands. Above the exceptional point [cf. Eq.~\eqref{opty23sdsdsdads23}], the final spectral term is defined using the auxiliary functions $S_+^{>} (\omega)$ and $S_-^{>} (\omega)$, which describe the twin satellite peaks helping to form the celebrated spectral triplet discovered by Mollow~\cite{Mollow1969}. These sidebands are peaked around the so-called Mollow frequency $\tilde{\Omega}$, and together with the central Lorentzian form the inelastic spectral response. At the exceptional point [cf. Eq.~\eqref{opty23sdsdsdads23a}] the final spectral term presents as an unshifted Student's $t$-distribution (with $\nu = 3$ degrees of freedom) such that the overall spectrum is a singlet. Furthermore, below the exceptional point [cf. Eq.~\eqref{opty23sdsdsdads23b}], the final term is the sum of two purely Lorentzian lineshapes (with different weighting coefficients $L_+^{<}$ and $L_-^{<}$). These Lorentzians are centered at resonance such that passing through the exceptional point leads to an interesting spectral transition from a simple singlet structure to an intriguing triplet. The exact forms of the auxiliary functions needed to fully describe the spectrum, as given in Eq.~\eqref{opty23sdsdsdads23} and Eq.~\eqref{opty23sdsdsdads23b}, are defined by
\begin{widetext} 
\begin{align}
\label{fdgfdhukjhjkhjkhjkg}
 S_\tau^{>} (\omega) &= \frac{1}{2\pi} \left[ \frac{1}{2} -  \frac{ \left( \frac{\gamma}{2} \right)^2}{2 \Omega^2 + \left( \frac{\gamma}{2} \right)^2} \right] \frac{ \tfrac{3\gamma}{4} }{ \left( \tfrac{3\gamma}{4} \right)^2 + \left( \omega - \omega_0 - \tau \tilde{\Omega} \right)^2 } - \frac{\tau}{16\pi} \frac{\gamma}{\tilde{\Omega}} \Biggr[ \frac{10\Omega^2- \left( \frac{\gamma}{2} \right)^2}{2\Omega^2 +  \left( \frac{\gamma}{2} \right)^2} \Biggr] \frac{  \omega - \omega_0 - \tau \tilde{\Omega}  }{ \left( \tfrac{3\gamma}{4} \right)^2 + \left( \omega - \omega_0 - \tau \tilde{\Omega} \right)^2 } , \\
  L_+^{<} &= \frac{\Omega^2}{\Gamma} \frac{ \frac{\gamma}{2} \left( \frac{\gamma}{2} + 2\Gamma \right) - 4\Omega^2 }{ \Bigr[ \left( \frac{\gamma}{2} \right)^2 + 2\Omega^2 \Bigr] \Bigr[ 2\Gamma - \frac{\gamma}{2}  \Bigr] }, 
  \quad\quad\quad\quad\quad
    L_-^{<} = \frac{8 \Omega^4}{ 32 \Omega^4 - \gamma \left( 10\Gamma + \frac{9\gamma}{2} \right) \Omega^2 + \left( 2 \Gamma + \frac{\gamma}{2} \right) \left( \frac{\gamma}{2} \right)^3 }. \label{fdgfdhukjhjkhjkhjkgB} 
\end{align}
\end{widetext} 
Equation~\eqref{fdgfdhukjhjkhjkhjkg} is composed of a weighted Lorentzian part (the first term) which is associated with spontaneous emission and a weighted dispersive part (the second term) due to interferences, which helps to define the Mollow triplet present for $\Omega > \Omega_{\mathrm{EP}}$. Notably, in the extreme limit of vanishing driving, the spectrum in Eq.~\eqref{opty23sdsdsdads23b} collapses into the solitary delta peak
\begin{equation}
\label{eqapp:sdfsdfsof_msdsdfdfsfsdsdfsdfdsdotion}
 \lim_{\Omega \to 0} S( \omega ) = \delta (\omega - \omega_0),
\end{equation}
a feature wholly coming from the Rayleigh scattering peak, which mathematically can be traced back to the steady state term in the $g^{(1)} (\tau)$ solutions as given in Eq.~\eqref{geetwo3ghjghj2sdsds3}. The evolution with $\Omega$ of the spectrum of the atom is plotted in Fig.~\ref{geeone}~(c). Most notably, the aforementioned Mollow triplet structure~\cite{Mollow1969} is striking for larger drivings (orange line), before a singlet spectrum emerges with weaker drivings near to the exceptional point case (dashed red line). Only a delta peak (dotted cyan line) survives with vanishing driving. These results demonstrate the critical significance of the exceptional point for the optical response of the atom, and the vivid consequences for the emission spectrum.
\\


\section{Conclusion}
\label{Sec:Conclusion}

Exceptional points are increasingly influential across modern physics, particularly in wave physics as manifested in mechanics~\cite{Berntson2013, Dolfo2018} and electromagnetics~\cite{Xu2020, Lakhtakia2020}. Their impact on quantum objects can be treated within an open systems approach, and here we have reconsidered perhaps the simplest example: that of a two-level atom. We have shown the consequences of passing through an exceptional point on the response of the atom, including striking changes to its optical spectrum (which evolves from being a singlet to a beautiful triplet) and significant reconstructions of correlation functions (which transition from being non-oscillatory to oscillatory). We believe that looking at quantum systems from the viewpoint of exceptional points is rather instructive, since it (i) neatly categorizes distinct behavioural regimes and (ii) helps to explain the character of common observables via the properties of complex eigenvalues arising from non-Hermitian matrices. This perspective may become increasingly widespread with the continuing rise in popularity of non-Hermitian quantum physics~\cite{Bender2023}. 
\\


\noindent \textbf{Author declarations}\\
The authors have no conflicts of interest to disclose.
\\

\noindent \textbf{Acknowledgments}\\
\textit{Funding}: CAD is supported by the Royal Society via a University Research Fellowship (URF\slash R1\slash 201158) and an International Exchanges grant (IES\slash R1\slash 241078) with A.~Cidrim (Universidade Federal de São Carlos, Brazil). VAS acknowledges support from the HORIZON EUROPE Marie Sklodowska-Curie Actions (2021-PF-01, Project No. 101065500, TeraExc).\\




\end{document}